\documentclass[12pt]{article}
\usepackage{amssymb,amsmath,epsfig}
\begin{document}
\parindent=0pt
\parskip=6pt
\rm

\begin{center}
{\bf \Large Diamagnetic susceptibility of spin-triplet
ferromagnetic superconductors}

\vspace{0.5cm}

{\bf H. Belich$^{1,2}$, Octavio D. Rodriguez Salmon$^{1}$, Diana
V. Shopova$^{3}$,\\ and Dimo I. Uzunov$^{1,3 \dag}$}

\vspace{0.5cm}

$^{1}${\em International Institute of Physics, Universidade
Federal de Rio Grande do Norte, av. Odilon Gomes de Lima, 1722,
59078--400, Natal (RN), Brazil.}

$^{2}$ {\em Universidade Federal do Esp\'{\i}rito Santo (UFES),
Departamento de F\'{\i}sica e Qu\'{\i}mica, Av. Fernando Ferrari
514, Vit\'{o}ria, ES, CEP 29075-910, Brazil.}

$^{3}$ {\em Collective Phenomena Laboratory, G. Nadjakov Institute
of Solid State Physics,\\
 Bulgarian Academy of Sciences, BG-1784 Sofia, Bulgaria.}
\end{center}

$^{\dag}$ Corresponding author: d.i.uzunov@gmail.com\\



\begin{abstract}
We calculate the diamagnetic susceptibility in zero external
magnetic field above the phase transition from  ferromagnetic
phase to phase of coexistence of ferromagnetic order and
unconventional superconductivity. For this aim we use
generalized Ginzburg-Landau free energy of unconventional
ferromagnetic superconductor with spin-triplet electron pairing. A
possible application of the result to some intermetallic compounds
is briefly discussed.
\end{abstract}

\normalsize

\vspace{0.5cm}

\begin{figure}
\begin{center}
\epsfig{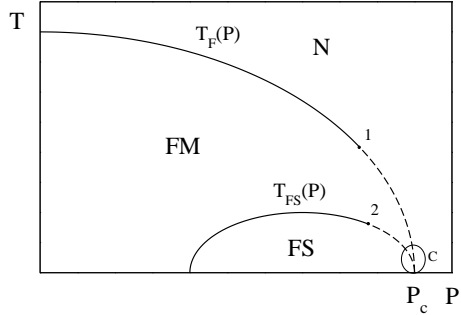}
\end{center}
\caption{\footnotesize An illustration of the $T-P$ phase diagram
of UGe$_2$ (details are omitted): N -- normal phase, FM -
ferromagnetic phase, FS - phase of coexistence of ferromagnetic
order and superconductivity, $T_{F}(P)$ and $T_{FS}(P)$ are the
respective phase transition lines (solid line corresponds to
second order phase transition, dashed lines correspond to first
order phase transitions; $1$ and $2$ are tricritical points;
$P_c\sim 1.6$ GPa is the critical pressure; $T_{F}(0) \sim 53$ K;
$T_{FS} < 1.22$ K; the loop C indicates a small domain ($T<0.3$ K,
$P\sim 16$ GPa) where the shape of the phase diagram is not well
established by available experimental data.} \label{Fig1}
\end{figure}

In certain ferromagnetic unconventional superconductors the phase
transition to superconductivity states occurs in the domain of
stability of ferromagnetic phase (an example is the itinerant
ferromagnet UGe$_2$~\cite{Saxena:2000,Huxley:2001,Tateiwa:2001}).
This seems to be a general feature of ferromagnetic
superconductors with spin-triplet electron
pairing~\cite{Shopova:2003, Cottam:2008, Shopova:2009} (see also
reviews~\cite{Shopova:2005, Shopova:2006}). In such situation the
thermodynamic properties near the phase transition line may differ
from those known for the superconducting-to-normal metal
transition. We show this by using the example of  diamagnetic
susceptibility above the phase transition line of superconducting
transition in spin-triplet ferromagnetic superconductors. This is
the line in the temperature-pressure ($T-P$) phase diagram
(Fig.~\ref{Fig1}), which separates the pure ferromagnetic phase
(FM) and the phase (FS) of coexistence of ferromagnetic order and
superconductivity. Here we present the result for diamagnetic
susceptibility which follows from the Ginzburg--Landau theory for
such type of superconductors~\cite{Shopova:2003, Cottam:2008,
Shopova:2009}. We outline the main steps of  calculation of
diamagnetic susceptibility in the Gaussian approximation. At the
end we briefly discuss the possible application of our results to
real systems.

Following notations and results in Refs.~\cite{Shopova:2003,
Shopova:2005, Shopova:2006}, we present the GL free energy
(fluctuation Hamiltonian) of spin-triplet ferromagnetic
superconductors, which is essential in the present consideration,
namely

\begin{equation}
\label{eq1} {\cal{H}} = \int d^3 x
\left\{\hat{H}_0\left[\psi(\mathbf{x})\right] +
\hat{H}_M\left[\psi(\mathbf{x})\right]\right\}\,
\end{equation}
\noindent by the energy densities

\begin{equation}
\label{eq2} \hat{H}_0 =  \frac{\hbar^2}{4m}\sum_{j=1}^{3} \left |
\left (\nabla -
 \frac{2ie}{\hbar c}\mathbf{A}\right )\psi_j \right|^2
+ a_s|\psi|^2
\end{equation}
\noindent
and

\begin{equation}
\label{eq3} \hat{H}_M = i\gamma_0 \mathbf{M}\cdot(\psi\times
\psi^*) + \rho\mathbf{M}^2\cdot\psi^2
\end{equation}
\noindent In Eqs.~(\ref{eq2})--(\ref{eq3}), $\psi(\mathbf{x}) =
\left\{\psi_j(\mathbf{x}); j = 1,2,3\right\}$ is three dimensional
vector field with complex components $\psi_j$, which represents
the superconducting order, $\mathbf{M}$ is the spontaneous
magnetization, the vector potential $\mathbf{A}$ is related to the
magnetic induction by $\mathbf{B} = \mathbf{H} + 4\pi\mathbf{M}$
and obeys the Coulomb gauge $(\nabla\cdot\mathbf{A}=0)$; $a_s =
\alpha_s(T-T_s)$, $\gamma_0$ and $\rho$ are positive material
parameters, and $2e$ and $2m$ are the charge and the effective
mass of the electron Cooper pairs, respectively. We neglect the
possible spatial anisotropy, which is usually represented in the
gradient terms of the Hamiltonian ${\cal{H}}$ (see, e.g.,
Ref.~\cite{Shopova:2003, Shopova:2005, Shopova:2006}).

Our task is to calculate the equilibrium free energy

\begin{equation}
\label{eq4} F = -\beta^{-1}\ln\int\prod_{\mathbf{x} \in
V}{\cal{D}}\psi(\mathbf{x})\exp{(-\beta \hat{H})},
\end{equation}
\noindent in the volume $V= L_xL_yL_z$ of the superconductor and
the diamagnetic susceptibility per unit volume in zero external
magnetic field, given by $\chi =
[-\partial^2F/V\partial^2H]_{H=0}$; $\beta^{-1} = k_BT$. In
Eq.~(\ref{eq4}), the functional integral is taken over both real
[$\Re\psi(\mathbf{x})$] and imaginary [$\Im\psi(\mathbf{x})$]
parts of the complex field $\psi(\mathbf{x})$, i.e.,
${\cal{D}}\psi(\mathbf{x}) \equiv
d\Re\psi(\mathbf{x})d\Im\psi(\mathbf{x})$. Note that for
temperatures near $T_{FS}(P)$ we can always set $\beta \approx
\beta_{FS}= 1/k_BT_{FS}$ (see, e.g.,~\cite{Uzunov:2010}).

As far as the behaviour in FM phase in a close vicinity of curve
$T_{FS}(P)$ is of interest to our consideration, (see
Fig.~\ref{Fig1}), the magnetization $\boldmath{M}$ has a magnitude
$|\mathbf{M}|\equiv M$, given by $M(T,P) =
[\alpha_f(T-T_F)/b_f]^{1/2}$, i.e., the result from the standard
Landau theory of ferromagnetic transitions with parameters $a_f =
\alpha_f(T-T_F)$ and $b_f$~\cite{Shopova:2003}

\begin{equation}
\label{eq5} F_m = a_{f}\mathbf{M}^2 + \frac{b_f}{2}\mathbf{M}^4,
\end{equation}
\noindent where $a_f = \alpha_{f}(T-T_F)$, and $b_f > 0$.
Therefore, in our consideration $M(T,P)$ is a known thermodynamic
quantity, which is established by the exhaustive thermodynamic
analysis of the phases in the unconventional superconductor
in~\cite{Shopova:2003}.

We choose the magnetization $\mathbf{M} = (0,0,M)$ and the
external magnetic field $\mathbf{H}=(0,0,H)$ to lie along the
$\hat{z}$-axis. Then the first term in Eq.~(\ref{eq3}) takes the
simple form $M(\psi\times\psi^{\ast})_z = M(\psi_1\psi_2^{\ast} -
c.c.)$. Under the supposition of uniform external magnetic field
$H$,  we take the gauge of the vector potential $\mathbf{A}$ as
$\mathbf{A} = (-By,0,0)$, and following classic
papers~\cite{Schmidt:1968, Schmid:1969, Pitaevskii:1980}, we can
represent the fields $\psi_j(\mathbf{x})$ by the series

\begin{equation}
\label{eq6} \psi_j(\mathbf{x}) =
\frac{1}{L_xL_z}\sum_{q}c_j(q)\varphi_j(q, \mathbf{x})\,
\end{equation}
\noindent in terms of the eigenfunctions

\begin{equation}
\label{eq7} \varphi_j(\mathbf{q}, \mathbf{x}) =
\frac{1}{\left(L_xL_z\right)^{1/2}}e^{i(k_x+k_z)}u_n(y)
\end{equation}
\noindent of the operator $\left[i\hbar\nabla +
(2e/c)\mathbf{A}\right]^2/4m$, corresponding to the eigenvalues

\begin{equation}
\label{eq8} E(q) =  \left(n + \frac{1}{2}\right)\hbar\omega_c +
\frac{\hbar^{2}}{4m}k_z^2,
\end{equation}
specified by the quantum number $n= 0, 1, \dots, \infty$, the wave
vector components $k_x$ and $k_z$, and the cyclotron frequency
$\omega_c = (eB/mc)$. In Eq.~(\ref{eq6}), the function $u_n(y)$ is
related to the Hermite polynomials $H_n(y)$ by

\begin{equation}
\label{eq9}u_n(y) = A_n
e^{-\frac{(y-y_0)^2}{2a_H^2}}H_n\left(\dfrac{y-y_0}{a_H}\right),
\end{equation}
\noindent where $A_n^{-1} =
(a_B2^nn!\sqrt{\pi})^{1/2}$~\cite{Abramowitz:1965}, $y_0 =
a_B^2k_x$, and $a_B = (\hbar c/2|e|B)^{1/2}$; $B = |\mathbf{B}|$.

Now the fluctuation Hamiltonian becomes ${\cal{H}} =
\sum_{q}\hat{{\cal{H}}}(q)$ with

\begin{align}
\label{eq10} \hat{{\cal{H}}}(q) & = \sum_j \tilde{E}(q)
c_j(q)c_j^{\ast}(q) \nonumber \\ &+ i\gamma_0M \left[
c_1(q)c_2^{\ast}(q) - \mbox{c.c.}\right],
\end{align}
\noindent where

\begin{equation}
\label{eq11}\tilde{E}(q) = E(q)+a_s+\rho M^2.
\end{equation}

Applying the unitary transformation,

\begin{subequations}
\begin{equation}
\label{eq12a} c_1(q) = \frac{i}{\sqrt{2}}\left[-\phi_{+}(q) +
\phi_{-}(q)\right]
\end{equation}
\begin{equation}
\label{eq12b} c_2(q) = \frac{1}{\sqrt{2}}\left[\phi_{+}(q) +
\phi_{-}(q)\right]
\end{equation}
\end{subequations}
\noindent renders the fluctuation Hamiltonian as a sum
of squares of field components $c_3(q)$, and $\phi_{\pm}(q)$,
and the free energy (\ref{eq4}) can be calculated as usual Gaussian
integrals over the same fields.

Following approximations, justified in
Ref.~\cite{Pitaevskii:1980}, we obtain the result

\begin{equation}
\label{eq13} \frac{F}{V} = \mu B^2\left(\frac{1}{a_{-}^{1/2}}
+\frac{1}{a_0^{1/2}} + \frac{1}{a_{+}^{1/2}} \right),
\end{equation}
\noindent where

\begin{equation}
\label{eq14} a_{\pm}(\gamma_{0}) = a_s + \rho M^2 \pm \gamma_0 M,
\end{equation}
\noindent $a_0 \equiv a_{\pm}(0)$, and $\mu = e^2k_BT/24\pi\hbar
c^2m^{1/2}$. Having in mind that $\partial /\partial H =
\partial/\partial B$,  the  fluctuation diamagnetic susceptibility
in Gaussian approximation takes the form
\begin{equation}
\label{eq15} \chi(T) = -2\mu \left(\frac{1}{a_{-}^{1/2}}
+\frac{1}{a_0^{1/2}} + \frac{1}{a_{+}^{1/2}} \right),
\end{equation}
\noindent
In contrast to  usual
superconductors~\cite{Pitaevskii:1980}, where the contribution to
the free energy from the diamagnetic currents is represented by a
single term, here we have three terms with labels $0$, and $\pm $
which exactly correspond to the contributions of the field
components $c_3$, and $\phi_{\pm}$, respectively.

Now one should use known results~\cite{Shopova:2003, Cottam:2008,
Shopova:2009,Shopova:2005, Shopova:2006} to analyze the
singularities of free energy in a close vicinity $(0<T-T_{FS} \ll
T_{FS})$ to the phase transition curve $T_{FS}(P)$ in the FM phase
$(T_F>T>T_{FS})$, where $M(T,P) = [\alpha_f(T_F-T)/b_f]^{1/2}$
and, for some real intermetallic compounds, for example, UGe$_2$,
the condition $(T_F-T_{FS})\gg (T-T_{FS})$ is satisfied. We shall
briefly discuss the behaviour of the free energy (\ref{eq13}) near
the left-hand part of the curve $T_{FS}(P)$, where the phase
transition FM-FS is of second order. For this case the critical
temperature $T_{FS}(P)$ is given in Refs.~\cite{Cottam:2008,
Shopova:2009}. In the present notations $T_{FS}(P)$ is defined by
the equation

\begin{equation}
\label{eq16} T_{FS} = T_s - \frac{\rho}{\alpha_s}\Delta  +
\frac{\gamma_0}{\alpha_s}\Delta^{1/2},
\end{equation}
\noindent where $\Delta \equiv [M(T_{FS})]^2=
\alpha_f(T_F-T_{FS})/b_f > 0$. Expanding $a_0(T)$, and
$a_{\pm}(\gamma_0,T)$ to first order in $(T-T_{FS})$, one may
easily check that $a_{-}(T_{FS})=0$ and

\begin{equation}
\label{eq17} a_{-}(T) \approx \tilde{a}_{-}(T-T_{FS}),
\end{equation}
\noindent where

\begin{equation}
\label{eq18} \tilde{a}_{-} = \alpha_s - \frac{\rho\alpha_f}{b_f} +
\frac{\gamma_0\alpha_f^{1/2}}{2\left[b_f\left(T_{F}-T_{FS}\right)\right]^{1/2}},
\end{equation}
\noindent while $a_{0}$ and $a_{+}$ remain positive at $T_{FS}$:
$a_{0}(T_{FS}) =\gamma_0\Delta^{1/2}$, and  $a_{+}(T_{FS})=
2a_{0}(T_{FS})$. Therefore, only one of all three fluctuation
diamagnetic contributions in Eqs.~(\ref{eq13}) and (\ref{eq15})
will generate singularity of the free energy and the typical
divergence of susceptibility. Keeping only the singular term in
Eq.~(\ref{eq13}), we obtain that in a close vicinity of line
$T_{FS}(P)$, where $a_{-} \ll \mbox{min}(a_0, a_+)$,

\begin{subequations}
\begin{equation}
\label{eq19a} \chi(T) =
\frac{\chi_0}{\left(T-T_{FS}\right)^{1/2}},
\end{equation}
\noindent $(T > T_{FS})$, where the scaling amplitude $\chi_0$ is
given by
\begin{equation}
\label{eq19b} \chi_0=-\frac{2\mu}{ \tilde{a}_{-}^{1/2}}.
\end{equation}
\end{subequations}
\noindent Note that $a_{\pm}(\gamma_0)
> 0$ is a condition for the stability of the FM phase and,
therefore, the quantity $\tilde{a}_{-}$ is always positive for
$T_{FS}(P) < T < T_{F}(P)$.

The formulae (\ref{eq19a}) and (\ref{eq19b}) are our main result.
This scaling relation~\cite{Uzunov:2010} is of typical Gaussian
type with an inverse root dependence on $(T-T_{FS})$ whereas the
scaling amplitude $\chi_0$ contains an essentially new
information. Compared to known result for usual
superconductors~\cite{Pitaevskii:1980}, the fluctuation
diamagnetic susceptibility (\ref{eq19a}) contains an extra factor
$(\tilde{a}_{-})^{-1/2}$, which depends on the material parameters
of the unconventional ferromagnetic superconductor. The value of
the new susceptibility amplitude factor $(\tilde{a}_{-})^{-1/2}$
in Eq.~(\ref{eq19b}) should be taken at $T_{FS}(P)$ for any
pressure $P$ of interest. Thus in evaluating the parameter
$\tilde{a}_{-}$ we may use the Eq.~(\ref{eq16}) for $T_{FS}(P)$.

In some real systems the Eq.~(\ref{eq18}) can be simplified. For
example, in UGe$_2$, $T_s \sim 0 $ K \cite{Cottam:2008,
Shopova:2009}, $T_{F} \gg T_{FS}$ \cite{Saxena:2000} and,
therefore, one may use $\tilde{a}_{-} \approx (\alpha_s -
\rho\alpha_f/2b_f)$. This result is obtained with the help of
Eq.~(\ref{eq16}). In itinerant ferromagnets with uniaxial
anisotropy as, for example, UGe$_2$, both phases FM and FS may
occur in two domains with opposite magnetizations $|\mathbf{M}|
=\pm M$. Here we have considered FM and FS with $M>0$. In the
domains of FM, where $M < 0$, the singular parts of the free
energy and the susceptibility will be given by the terms,
containing the quantity $a_{+}$. Because of the invariance of the
Eqs.~(\ref{eq13}) and (\ref{eq15}) with respect to the change
$a_{\pm} \rightarrow a_{\mp}$, the results presented by
Eqs.~(\ref{eq13}), (\ref{eq15}), and (\ref{eq19a})--(\ref{eq19b})
are valid in both domains of the FM and $\psi$-fluctuations
corresponding to any domain ($M\lessgtr 0$) of
FS~\cite{Shopova:2003}.

We have used the Gaussian approximation, which is not valid in the
critical region~\cite{Uzunov:2010} of anomalous fluctuations.
However, the critical region of real ferromagnetic superconductors
with spin-triplet electron pairing is often very narrow and,
hence, virtually of no interest. Therefore, the present results
can be reliably used in interpretation of experimental data for
real itinerant ferromagnets, which exhibit low-temperature
spin-triplet superconductivity triggered by the ferromagnetic
order.

 \end{document}